\documentclass[prd,twocolumn,showpacs,preprintnumbers,amsmath,amssymb,floatfix,nofootinbib]{revtex4}

\usepackage{subfigure,graphicx,epsfig,amsmath,amsfonts,amssymb,xcolor,slashed,ulem}

\usepackage{bm}
\usepackage{graphicx}
\usepackage{amsmath}
\usepackage{amsfonts}
\usepackage{amssymb}
\usepackage{color}
\usepackage{ulem}

\usepackage[section]{placeins}

\usepackage{booktabs}
\usepackage[colorlinks, citecolor=blue,anchorcolor=red,menucolor=red, linkcolor=red,filecolor=red,runcolor=red,urlcolor=blue,frenchlinks=red]{hyperref}

\newcommand{\be}{\begin{equation}}
\newcommand{\ee}{\end{equation}}
\newcommand{\ba}{\begin{eqnarray}}
\newcommand{\ea}{\end{eqnarray}}
\newcommand{\nn}{\nonumber}

\newcommand{\mev}{\textrm{ MeV}}

\begin{document}
\bibliographystyle{unsrt}
\arraycolsep1.5pt

\title {Signatures of the two $K_1(1270)$ poles in $D^+\to \nu e^+ V P$ decay}

\author{Guan-Ying Wang}

\affiliation{School of Physics and Microelectronics, Zhengzhou University, Zhengzhou, Henan, 450001, China
}

\author{Luis Roca}
\affiliation{\it  Departamento de F\'{\i}sica. Universidad de
Murcia. E-30100 Murcia.  Spain.}

\author{En Wang}
\affiliation{School of Physics and Microelectronics, Zhengzhou University, Zhengzhou, Henan 450001, China}

\author{Wei-Hong Liang}
\affiliation{Department of Physics, Guangxi Normal University, Guilin 541004, China}
\affiliation{Guangxi Key Laboratory of Nuclear Physics and Technology, Guangxi Normal University, Guilin 541004, China}

\author{Eulogio Oset}

\affiliation{Departamento de F\'{\i}sica Te\'orica e IFIC, Centro Mixto
Universidad de Valencia-CSIC, Institutos de Investigacion de
Paterna, Apdo 22085, 46071 Valencia, Spain}

\begin{abstract}

We analyze theoretically
the  $D^+\to \nu e^+ \rho \bar K$ and $D^+\to \nu e^+ \bar K^* \pi$ decays to see the feasibility to check the double pole nature of the  axial-vector resonance $K_1(1270)$ predicted by the unitary extensions of  chiral perturbation theory (UChPT). Indeed, 
within UChPT the $K_1(1270)$ is dynamically generated from the interaction of a vector and a pseudoscalar meson,  and two poles are obtained for the quantum numbers of this resonance. The lower mass pole couples dominantly to $K^*\pi$ and the higher mass pole to $\rho K$, therefore we can expect that different reactions weighing differently these channels in the production mechanisms enhance one or the other pole. 
We show that the different final $VP$ channels in $D^+\to \nu e^+ V P$ weigh differently both poles, and this is reflected in the shape of the final vector-pseudoscalar invariant mass distributions. Therefore, we conclude that
these decays are suitable to distinguish experimentally the 
predicted double pole of the $K_1(1270)$  resonance.

\end{abstract}

\maketitle

\section{Introduction}

Semileptonic $B$ and $D$ meson decays have been for long considered as a good source to learn about non perturbative strong interactions, given the good knowledge of the weak vertex \cite{Isgur:1988gb,Scora:1995ty,Bajc:1995km}. Refined methods have become available more recently \cite{Soni:2018adu,Yao:2018tqn} and the reactions are looked upon with interest to even learn about physics beyond the standard model \cite{Fajfer:2012vx,Tanaka:2012nw}. Explicit calculations correlating a vast amount of data with the help of some selected pieces of experimental information are also available \cite{Dai:2018vzz}. 
One of the relevant cases of these reactions consist on $D$ meson decays leading to resonances in the final state, rather than the ordinary ground state of mesons, usually studied. 
   In particular, semileptonic decays of hadrons where the final hadron is a resonance are specially interesting. 
In this sense,  the $B$ and $B_s$ semileptonic decays leading to $D_0^*(2400)$ and $D_{s0}^*(2317)$ resonances were studied in Ref.~\cite{Navarra:2015iea}. Similarly, the $D$ decays into the scalar mesons $f_0(500)$, $K_0^*(800)$, $f_0(980)$ and 
$a_0(980)$ were addressed in Ref.~\cite{Sekihara:2015iha}, with relevant results concerning the nature of these scalar mesons. A review of these and related reactions can be seen in Ref.~\cite{Oset:2016lyh}.
In this direction, the recent observation of the $D^+\to \nu e^+ \bar K_1^0(1270)$ reaction measured by the BESIII collaboration 
\cite{Ablikim:2019wxs} offers a new opportunity to study the properties and nature of the $K_1(1270)$ axial-vector resonance. Prior to this measurement the CLEO collaboration presented results on the $D^+\to \nu e^+ \bar K_1^0(1270)$ \cite{Artuso:2007wa}, but the quality of data is much improved in the BESIII measurements. Interestingly there are theoretical results on these reactions in Refs.~\cite{Isgur:1988gb,Scora:1995ty} using quark models, in Ref.~\cite{Khosravi:2008jw} using QCD sum rules and factorization approach, in Ref.~\cite{Cheng:2017pcq} using a covariant light front quark model and in 
Ref.~\cite{Momeni:2019uag} using light cone sum rules. The branching ratios obtained, within $10^{-2}-10^{-3}$, agree qualitatively with the one measured by BESIII of about $2.3\times 10^{-3}$.

Our interest in this reaction stems from the findings of Refs.~\cite{Roca:2005nm,Geng:2006yb} that there are two $K_1(1270)$ resonances instead of one. The idea of the present work is to see which are the particular measurements in the BESIII reaction that could show evidence of these two states, for which we do theoretical calculations looking into particular final channels.
 The standard quark model picture for mesons and baryons
\cite{isgurmeson,isgurbaryon,capstick,roberts,vijande}
has the great value to correlate a great amount of data on hadron spectroscopy, but the axial vector meson states are systematically not so well reproduced as the vector ones \cite{isgurmeson,vijande}. With this perspective it is not surprising that other pictures have been proposed to explain these states. The chiral unitary approach 
\cite{Kaiser:1995cy,Oller:1997ti} was applied to the study of the pseudoscalar-vector meson interaction, using the chiral Lagrangian of Ref.~\cite{Birse:1996hd} and it was shown that the interaction, unitarized in coupled channels, gave rise to bound states or resonances which could be identified with  the low lying axial-vector resonances
\cite{lutz,Roca:2005nm,chengeng}. An appealing feature of these dynamically generated resonances is that the reaction mechanisms producing them proceed in a different way than ordinary mechanisms that produce resonances. Indeed, one does not produce the resonances directly, rather one produces the meson-meson components of the different coupled channels, which upon final state interaction among themselves generate the resonances. This allows one to perform calculations and relate many production channels, and often leads to particular features in the invariant mass distributions \cite{Oset:2016lyh}.
Concerning axial-vector meson production in different reactions, work has been done recently in the study of the $J/\psi\to \eta(\eta')h_1(1380)$ reaction \cite{liangsakai},
$\tau^-\to\nu_\tau P A$ with $P=\pi,K$ and $A=$ $b_1(1235)$, $h_1(1170)$, $h_1(1380)$, $a_1(1260)$, $f_1(1285)$
\cite{dairoca} and $\chi_{cJ}$ decay to $\phi h_1(1380)$
\cite{Jiang:2019ijx}, among others quoted in those works.

In Refs.~\cite{Roca:2005nm,Geng:2006yb} it was shown that there were two $K_1(1270)$ states, which coupled differently to the coupled channels. One state appears at $1195\mev$ and couples mostly to $K^*\pi$. The other state appears at $1284\mev$ coupling mostly to $\rho K$.
In Ref.~\cite{Geng:2006yb} some reactions disclosing these final states were studied and it was shown that they peaked at different energies, and the state of higher mass had a smaller width. The existence of the two $K_1(1270)$ is directly linked to the chiral dynamics of the problem and is similar to the appearance of the two $\Lambda(1405)$ states in the baryon strange sector
\cite{Oller:2000fj,Jido:2003cb,Hyodo:2011ur} (see the review ``{\it Pole structure of the $\Lambda(1405)$ region}"  in  the PDG~\cite{pdg}). With this picture in mind some reactions have been proposed to provide extra evidence of the existence of these two $K_1(1270)$ states. In Ref.~\cite{dairoca} the $\tau^-\to\nu_\tau K^- K_1(1270)$ reaction is proposed looking at the $\rho K$ and $K^* \pi$ final decay products of the $K_1(1270)$ and two distinct peaks are seen in the results. In Ref.~\cite{Wang:2019mph} the $D^0 \to \pi^+ \rho \bar K$ and $D^0 \to \pi^+ \pi \bar K^*$ reactions are also suggested in order to see the two peaks  corresponding to the two $K_1(1270)$ resonances.

In the present work, taking advantage of the recent BESIII measurement \cite{Ablikim:2019wxs}, we look at the 
$D^+\to \nu e^+ \bar K_1^0(1270)$ reaction, evaluating explicitly the decays  $\bar K_1(1270)\to \rho \bar K$ and $\bar K_1(1270)\to \pi \bar K^*$
showing that these final channels give different weights to the two $K_1(1270)$ resonances and lead to invariant mass distributions that differ in the position and the shape.
In view of the results obtained here we can only encourage the BESIII collaboration to perform the analysis that we suggest here, which should shed valuable light on the issue of the two $K_1(1270)$ states and the nature of the low lying axial-vector resonances.

\section{Formalism}


\begin{figure}[h]
\begin{center}
\includegraphics[width=0.5\textwidth]{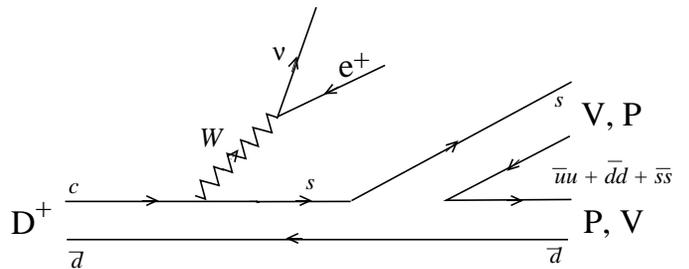}
\caption{\small{Elementary $D^+\to \nu e^+ V P$ process at the quark level.}}
\label{fig:Tquarks}
\end{center}
\end{figure}

As explained in the introduction, within the chiral unitary approach (UChPT) of Refs.~\cite{Roca:2005nm,Geng:2006yb}, the axial-vector resonances  are generated dynamically by the non-linear chiral dynamics involved in the unitarization procedure of the elementary $VP$ scattering potential in s-wave, and there is no need to include them as explicit degrees of freedom (by means of Breit-Wigner like amplitudes or similar). (We refer to \cite{Roca:2005nm} for the seminal work on the UChPT approach for the axial-vector resonances, and to \cite{Geng:2006yb,Wang:2019mph,Geng:2008ag,Roca:2012rx} for brief but illustrative summaries).
In particular, for the strangeness $S=1$ and isospin $I=1/2$ channels two poles were found in Refs.~\cite{Roca:2005nm,Geng:2006yb}, which were associated to two $K_1(1270)$ resonances, looking at unphysical Riemann sheets of the unitarized $VP$ scattering amplitudes. The poles are located  at $(1195-i123)\mev$ and $(1284-i73)\mev$, where we can identify the real part with the mass and the imaginary part with half the width. In Table IV of Ref.~\cite{Geng:2006yb} the values of the different couplings to the different $VP$ channels can be seen. The main observation is that the lower mass pole couples dominantly to $K^*\pi$ and the higher mass pole to $\rho K$, but the couplings to the other $VP$ channels are not negligible, and are actually considered. 
Following this philosophy, the way to produce a dynamically generated $K_1(1270)$ resonance in a particular reaction is to create first all possible $VP$ pairs and then implement their final state interaction. 
 This later issue will be addressed in the second part of this section but first we need to discuss the calculation of the 
elementary production of the $VP$ states, and its depiction, at the quark level,  can be seen in Fig.~\ref{fig:Tquarks}.
First the $c$ quark produces an $s$ quark through the Cabibbo favored vertex $Wcs$ and then hadronization into a final vector and a pseudoscalar meson is implemented by producing an extra $\bar{q} q$ with  the $^3P_0$ model  \cite{micu,LeYaouanc:1972vsx,bijker}.

We are mostly interested in evaluating the relative weight and momentum dependence of the different channels modulo a global arbitrary normalization factor.
The different weights among the allowed $VP$ channels can be obtained from the following $SU(3)$ reasoning.

The flavor state  of the final hadronic part after the $\bar{q} q$ is produced in the hadronization is
\begin{align*}
|H\rangle \equiv |s\,(\bar u u +\bar d d +\bar s s)\,\bar d \rangle,
\end{align*}
which can be written as 
\begin{align}
|H\rangle = \sum_{i=1}^3{|M_{3i}q_i\bar d} \rangle=
            \sum_{i=1}^3{|M_{3i}M_{i2} }    \rangle=
	     |(M^2)_{32}\rangle,
\label{eq:Hket}
\end{align}
where we have defined

\begin{equation}
q\equiv \left(\begin{array}{c}u\\d\\s\end{array}\right)\,\text{~~and~~~}
M\equiv q\bar q^\intercal=\left(\begin{array}{ccc}u\bar u & u\bar d & u\bar s\\
					       d\bar u & d\bar d & d\bar s\\
					       s\bar u & s\bar d & s\bar s
\end{array}\right)\,.
\label{eq:Mqqbar}
\end{equation} 

The hadronic states can be identified with the physical mesons associating the $M$ matrix with the usual $SU(3)$ matrices containing the pseudoscalar and vector mesons:

\begin{align*}
M\Rightarrow P\equiv
\left(\begin{array}{ccc} 
              \frac{\pi^0}{\sqrt{2}}  + \frac{\eta}{\sqrt{3}}+\frac{\eta'}{\sqrt{6}}& \pi^+ & K^+\\
              \pi^-& -\frac{1}{\sqrt{2}}\pi^0 + \frac{\eta}{\sqrt{3}}+ \frac{\eta'}{\sqrt{6}}& K^0\\
              K^-& \bar{K}^0 & -\frac{\eta}{\sqrt{3}}+ \frac{2\eta'}{\sqrt{6}} 
      \end{array}
\right)\,,
\label{eq:Pmatrix}
\end{align*}

\begin{equation}
 M\Rightarrow V \equiv
 \left(\begin{array}{ccc} \frac{1}{\sqrt{2}} \rho^0 +
\frac{1}{\sqrt{2}}\omega
 & \rho^+ & K^{*+}\\
\rho^-& - \frac{1}{\sqrt{2}} \rho^0 + \frac{1}{\sqrt{2}}\omega
& K^{*0}\\
K^{*-}& \bar{K}^{*0} & \phi
\end{array}
\right), \label{eq:Vmatrix}
\end{equation}
where the usual mixing between the singlet and octet to give $\eta$ and $\eta'$ \cite{Bramon:1992kr} has been used in $P$. Also in the $V$ matrix,
ideal $\omega_1$-$\omega_8$ mixing has been considered to produce $\omega$ and $\phi$, to agree with the quark content of $M$ in Eq.~(\ref{eq:Mqqbar}).

Since the $M^2$ in Eq.~\eqref{eq:Hket} can refer either to $VP$ or $PV$, we need to evaluate the contribution
\begin{align}
(VP)_{32}+(PV)_{32}&=
\rho^+ K^-  - \frac{1}{\sqrt{2}} \rho^0 \bar{K}^{0}
+  {K^*}^-\pi^+  \nn \\
&- \frac{1}{\sqrt{2}} \bar{K}^{*0} \pi^0
+ \frac{1}{\sqrt{2}}\omega \bar{K}^{0}
+ \phi  \bar{K}^{0}
\label{eq:weightsVP}
\end{align}
where we see that the $\bar{K^*}^0\eta$ channel has been cancelled mathematically and the $\eta'$ is neglected   because of its large mass as done in the original work of the $VP$ interaction that generated the axial-vector $K_1(1270)$ \cite{Roca:2005nm}. The numerical coefficients in Eq.~\eqref{eq:weightsVP} in front of each $VP$ channel provide the relative strength of the different $VP$ channels.

The momentum structure of the amplitude corresponding to the mechanism in Fig.~\ref{fig:Tquarks} can be evaluated in a similar way to 
what was done in Refs.~\cite{Navarra:2015iea,Sekihara:2015iha}.
Indeed, the amplitude, $T$, for the process of Fig.~\ref{fig:Tquarks} can be factorized into the weak part and the hadronization part, and then it will be proportional to

\begin{align}
 L^\mu Q_\nu\, V_{\rm Had}
\label{eq:TLQV}
\end{align}
where global constant factors are omitted since we will perform the calculations up to a global normalization. In Eq.~\eqref{eq:TLQV}
$L^\mu=\bar u_\nu \gamma^\mu(1-\gamma_5)v_l$ is the leptonic current and $Q_\mu=\bar u_s \gamma_\mu(1-\gamma_5)u_c$ the quark current. The hadronization part $V_{\rm Had}$ will be discussed later on.

When evaluating the $D$ decay width of this process, we will need to square the amplitude and sum over the quark polarizations which gives
(see Ref.~\cite{Sekihara:2015iha} for explicit details and calculation)

\begin{equation}
\frac{1}{2} \sum _{\rm pol} |T|^{2}
= \frac{4 |V_{\rm had} |^{2}}{m_{l} m_{\nu} m_{D} M_{\rm{VP}}}
( p_{l} \cdot p_{D} ) ( p_{\nu} \cdot p_{\rm{VP}} ) .
\label{eq:amp2}
\end{equation}
where $p_i$ are the four-momenta of the corresponding particles, $m_i$ the masses, and  the $\rm{VP}$ label refers to the final $VP$ pair, which will eventually account for  the $K_1(1270)$ resonance.

The final expression for the $VP$ invariant mass, $M_{\rm{VP}}$, distribution of the $D^+\to \nu e^+ V P$ decay can be obtained in the same  way as in Ref.\cite{Sekihara:2015iha} (see the derivation leading to Eq.(23) of Ref.~\cite{Sekihara:2015iha}) and gives

\begin{align}
\frac{d \Gamma}{d M_{\rm{VP}}} 
= \frac{2 }
{(2\pi)^{5}m_{D}^{3} M_{\rm{VP}}} 
&\int  d M_{e\nu } \, M_{e\nu}^2\,|\bm{p}_{\rm{VP}}|\, |\bm{\tilde{p}}_{\nu}| \,|\bm{\tilde{p}}_{V}|\nn \\
& \times\left ( \tilde{E}_{D} \tilde{E}_{\rm{VP}} - \frac{1}{3} | \tilde{\bm{p}}_{D} |^{2} 
\right )
|V_{\rm Had}|^2
\end{align}
where  $M_{e\nu }$ is the $e\nu$ invariant mass and 

\begin{align}
|\bm{p}_{\rm{VP}}|&=\frac{1}{2 m_D} \lambda^{1/2}(m_D^2, M_{e\nu}^2,M_{\rm{VP}}^2) \theta(m_D-M_{e\nu}-M_{\rm{VP}}), \nn \\
|\bm{\tilde{p}}_V|&=\frac{1}{2 M_{\rm{VP}}} \lambda^{1/2}(M_{\rm{VP}}^2, m_V^2,m_P^2) \theta(M_{\rm{VP}}-m_V-m_P), \nn \\
|\bm{\tilde{p}}_\nu|&=\frac{M_{e\nu}}{2},\nn \\
\tilde{E}_{D}&=\frac{m_D^2+M_{e\nu}^2-M_{\rm{VP}}^2}{2 M_{e\nu}}, \nn \\
\tilde{E}_{R}&=\frac{m_D^2-M_{e\nu}^2-M_{\rm{VP}}^2}{2 M_{e\nu}}, 
\end{align}
with $\lambda$ and $\theta$ standing for the  K\"all\'en and step functions respectively and we have neglected the positron mass.

\begin{figure}[h]
\begin{center}
\includegraphics[width=0.45\textwidth]{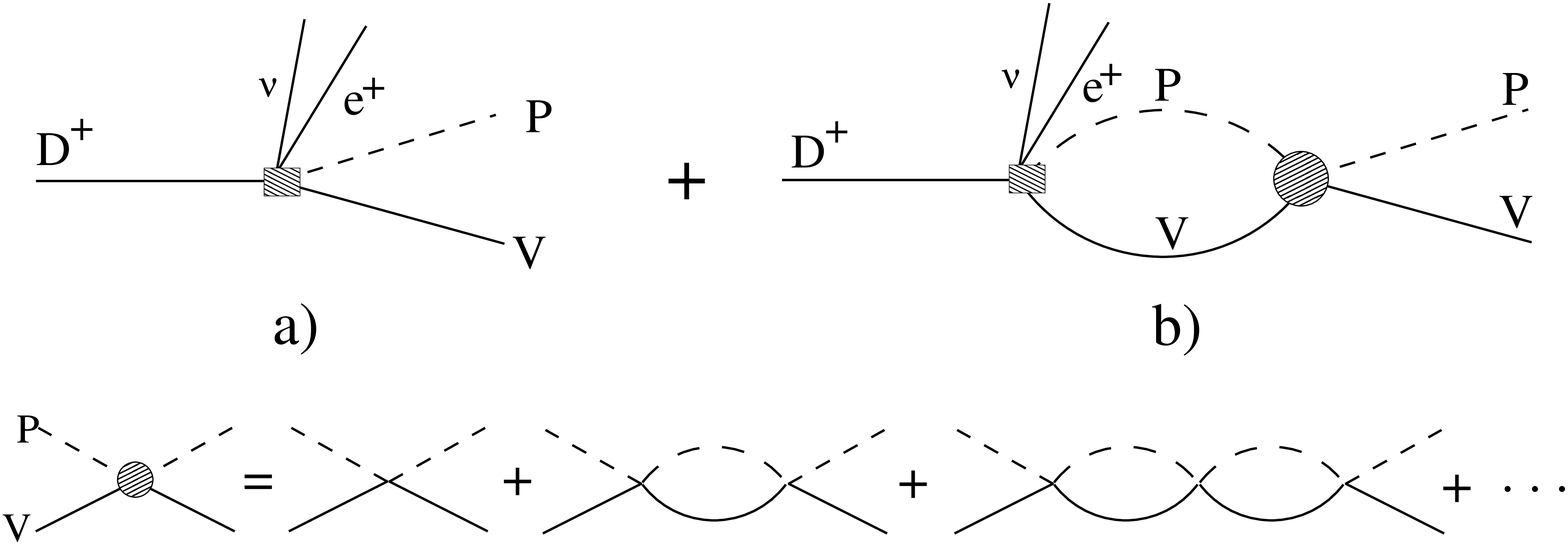}
\caption{\small{$VP$ final state interaction.}}
\label{fig:FSI}
\end{center}
\end{figure}

One of the main ingredients in the calculation of the hadronic part is the implementation of the final state interaction
 of the $VP$ pairs produced in the mechanism of Fig.~\ref{fig:Tquarks}, which is depicted in Fig.~\ref{fig:FSI}. Note that, since the $K_1(1270)$ resonance is generated dynamically within our approach, it is not  produced directly but, instead, the different $VP$ pairs are produced and then rescatter infinitely many times which is accounted for by the unitarized $VP$ scattering amplitude.

Taking into account the six different possible intermediate $VP$ pairs, (${K^*}^-\pi^+$, $\rho^+ K^-$, $\bar{K}^{*0} \pi^0$,  $\rho^0 \bar{K}^{0}$, $\omega \bar{K}^{0}$ and $\phi  \bar{K}^{0}$) the hadronic part of the amplitude for the decay into the $i-$th final $VP$ channel can be written as

\begin{equation}
V_{\rm Had}(D^+\to \nu e^+ V_iP_i)=V_p \left( h_i +
\sum_{j=1}^6 h_j G_j T^{\rm{I=1/2}}_{j,i}\right)
\label{eq:Vhadunit}
\end{equation}
where $V_p$ is an arbitrary global normalization factor,
which includes the weak coupling constant among other factors stemming from the quark matrix elements, 
 $h_i$ are the numerical coefficients in front of each $VP$ state in Eq.~\eqref{eq:weightsVP}, $G_j$ is the vector-pseudoscalar loop function \cite{Geng:2006yb} and $T^{\rm{I=1/2}}_{j,i}$ are the unitarized $(VP)_j\to(VP)_i$ scattering amplitude in isospin 1/2 from Ref.~\cite{Geng:2006yb}.
These are the amplitudes that manifest the double pole structure in the complex energy plane associated to the $K_1(1270)$.
Note that in Ref.~\cite{Geng:2006yb} the $VP$ states are in isospin basis and here we are working with explicit charge basis, but we can easily transform from one to the other basis using that
\begin{align}
| \rho \bar K \rangle_{I=\frac{1}{2}, I_3=\frac{1}{2}}&= \sqrt{\frac{2}{3}} |\rho^+ K^-\rangle - \frac{1}{\sqrt{3}} |\rho^0 \bar K^0\rangle,\nn\\
|\bar{K}^{*} \pi \rangle_{I=\frac{1}{2}, I_3=\frac{1}{2}}&= -\sqrt{\frac{2}{3}} |K^{*-} \pi^+\rangle + \frac{1}{\sqrt{3}} |\bar{K}^{*0}\pi^0\rangle.
\label{eq:changebasis}
\end{align}

 Note that these unitarized $VP$ scattering amplitudes do  not necessarilly have a Breit-Wigner shape in the real axis (see explicit plots in Refs.~\cite{Geng:2006yb,Wang:2019mph}). They actually contain the information of the whole $VP$ dynamics and not only the resonant structure. However, in a actual experiment one would typically try to fit Breit-Wigner like shapes and therefore we will also compare in the results section the results using for the scattering amplitudes 
\begin{equation}
T_{ij}=\frac{g_i g_j}{s-s_p},
\label{eq:tijpoles}
\end{equation}
where $s_p$ is the pole position which can be identified with the mass and width of the generated resonances $\sqrt{s_p}\simeq M_R -i \Gamma_R/2$ and $g_i$ are the couplings of the resonance to the $i-th$ $VP$ channel which can be obtained from the residues of the amplitudes at the pole positions and can be found in  Table IV of Ref.~\cite{Geng:2006yb}.

\section{Results}

\begin{figure*}[h]
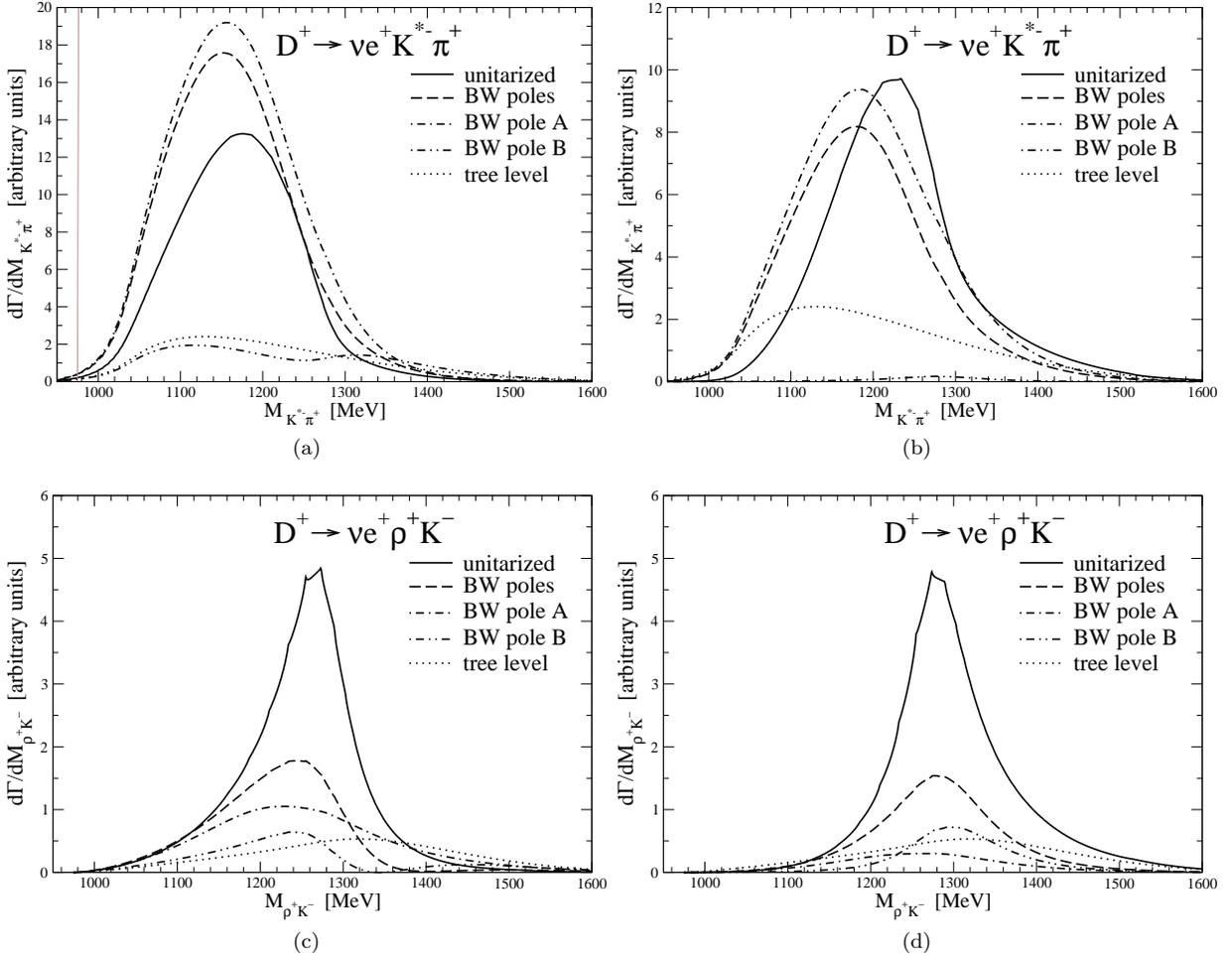

     \centering
     \subfigure[]{\includegraphics[width=.45\linewidth]{Minv1a.eps}}
     \subfigure[]{\includegraphics[width=.45\linewidth]{Minv1b.eps}} \\
     \subfigure[]{\includegraphics[width=.45\linewidth]{Minv2a.eps}}
     \subfigure[]{\includegraphics[width=.45\linewidth]{Minv2b.eps}} \\
   \caption{\small{$VP$ invariant mass distributions for  $D^+\to \nu e^+ {K}^{*-}\pi^+$ and $D^+\to \nu e^+ {\rho}^{+} K^-$. Left panels: including the interaction with the tree level mechanism  of Fig.~\ref{fig:Tquarks}. Right panels: without the interference with the tree level contribution.
}}
\label{fig:results1}
\end{figure*}

We first show in the left panels of Fig.~\ref{fig:results1} the different contributions to the $VP$ invariant mass distribution for the $D^+\to \nu e^+ {K}^{*-}\pi^+$ and  $D^+\to \nu e^+ {\rho}^{+} K^-$. The absolute normalization is arbitrary, but the relative strength between the different curves and the different channels are absolute (There is only a global normalization constant, the same for all the channels, see Eq.~\eqref{eq:Vhadunit}).
The label ``unitarized'' stands for the results using for the $VP\to VP$ amplitudes, $T^{\rm{I=1/2}}_{ij}$, the unitarized model from Ref.~\cite{Geng:2006yb}, as explained above. These curves are compared to the results using, instead, the explicit Breit-Wigner like shapes of Eq.\eqref{eq:tijpoles}, labeled as ``BW poles'' and also considering the contribution of only the lower mass pole  ($A$) or the higher mass pole ($B$).
The ``tree level'' curve represents the result removing the final $VP$ state interaction, {\it i.e.} only the mechanism of Fig.~\ref{fig:FSI}a), which is accounted for by considering only the  first $h_i$ term in Eq.~\eqref{eq:Vhadunit}.
We have also implemented a convolution with the final vector meson spectral function, in the same way as in  Ref.~\cite{Wang:2019mph}, in order to take into account the final vector meson widths. This is specially relevant  for the $\rho \bar K$ case due to the large width of the $\rho$ meson and the fact that the $\rho K$ threshold lies around the  $K_1(1270)$ energy region.

We see that the invariant mass distributions in these $D^+$ decays are clearly dominated by the $K_1(1270)$ resonant contribution but the curves are clearly different in shape and position of the peaks for the two final channels considered. 
Actually in the ${K}^{*-}\pi^+$ channel the peak of the distribution is located around 1160-1180~MeV, depending whether we use the unitarized or the Breit-Wigner amplitudes for the $VP$ scattering. 
However, for the ${\rho}^{+} K^-$ distribution the curve peaks at around 1250-1270~MeV  and is considerable narrower.
This is a clear manifestation of the different weight that the two $K_1(1270)$ poles have in both channels. Indeed, for the ${K}^{*-}\pi^+$ final channel, the distribution is clearly dominated by the lower mass pole, the one at $\sqrt{s_p}=(1195-i123)\mev$. This is a consequence of the large coupling of this pole to $\bar{K}^{*}\pi$. In the ${\rho}^{+}K^-$ channel the individual poles have a more comparable strength among themselves but the higher mass pole, the one at $(1284-i73)\mev$, shifts the final strength to higher energies and narrows the distribution.

It is also worth noting, however, that there is an important  interference effect between different mechanisms, particularly with the tree level contribution. This is clearly seen by comparing to the right panels, which have been evaluated removing the tree level terms, {\it i.e.} considering only the mechanisms in Fig.~\ref{fig:FSI}b). This is what one would obtain if 
the background, non-resonant terms could be ideally removed. In this later case the distributions would more clearly manifest the effect of the individual poles.

\section{Summary}

We show theoretically that the semileptonic decays of the $D^+$ meson into $\nu e^+ {K}^{*-}\pi^+$ and $\nu e^+ {\rho}^{+} K^-$ allows to distinguish the two different poles associated to the $K_1(1270)$ resonance as predicted by the chiral unitary   approach \cite{Roca:2005nm,Geng:2006yb}. 
Using as the only input the lowest order chiral perturbation theory Lagrangian accounting for the tree level interaction of a vector and a pseudoscalar meson, the implementation of unitarity in coupled channels allows to obtain the full $VP$ scattering amplitude which dynamically develops two poles associated to the $K_1(1270)$ resonance, without including them as explicit degrees of freedom. The poles show up naturally from the highly non-linear dynamics implied in the unitarization. 
Each pole has different features which could allow them to be distinguished in specifically devoted reactions, like those considered in the present work. Indeed, each pole couples differently to different $VP$ channels: the lower mass pole is wider and couples mostly to $K^*\pi$ and the higher mass pole is narrower and couples predominantly to $\rho K$.

The semileptonic decays studied in the present work proceed first with the elementary $VP$ production from the hadronization after the weak decay of the $c$ quark via the creation of a $q\bar q$ pair with  the $^3P_0$ model. The weight of the different channels are then related using $SU(3)$ arguments. The $K_1(1270)$ shows up in the decay after the implementation of the final state interaction of the $VP$ pair, using the unitarized $VP$ amplitudes. In spite of the fact that in the full amplitudes there is always a mixture of both poles, we obtain, by evaluating the $VP$ invariant mass distributions, that the $D^+\to \nu e^+ {K}^{*-}\pi^+$ weighs more the lower mass pole while in the 
$D^+\to \nu e^+ {\rho}^{+} K^-$ decay the higher mass pole has a greater influence. The shapes do not necessarilly reflect directly the pure resonant shape of each pole since there are interferences between the poles and non-resonant terms, but both the position and shape of the invariant mass distributions are clearly different and reflect the dominance of either pole in both channels considered and could be observed in experiments amenable to look at these mass distributions.

\section{Acknowledgments}
We would like  to acknowlege the fruitful discussions with Ju-Jun Xie and Li-Sheng Geng. This work is partly supported by the National Natural Science Foundation of China under Grant Nos. 11505158, 11847217, 11975083 and 11947413.  It is also supported by the Academic Improvement Project of Zhengzhou University.
This work is partly supported by the Spanish Ministerio
de Economia y Competitividad and European FEDER funds under Contracts No. FIS2017-84038-C2-1-P B
and No. FIS2017-84038-C2-2-P B.

\end{document}